# Chiral Dirac fermion in a collinear antiferromagnet

Ao Zhang(张奥)[1†], Ke Deng(邓可)[1†], Jieming Sheng(盛洁明)[1,2†], Pengfei Liu(刘鹏飞)[1†], Shiv Kumar[3], Kenya Shimada[3], Zhicheng Jiang(江志诚)[4], Zhengtai Liu(刘正太)[4], Dawei Shen(沈大伟)[4], Jiayu Li(李嘉裕)[1], Jun Ren(任俊)[1], Le Wang(王乐)[1], Liang Zhou(周良)[1], Yoshihisa Ishikawa[5], Takashi Ohhara[6], Qiang Zhang[7], Garry McIntyre[8], Dehong Yu[8], Enke Liu(刘恩克)[9], Liusuo Wu(吴留锁)[1*], Chaoyu Chen(陈朝宇)[1*] and Qihang Liu(刘奇航)[1*]

[1]Shenzhen Institute for Quantum Science and Technology and Department of Physics, Southern University of Science and Technology, Shenzhen, 518055, China

[2]Academy for Advanced Interdisciplinary Studies, Southern University of Science and Technology, Shenzhen, 518055, China

[3]Hiroshima Synchrotron Radiation Centre, Hiroshima University, Higashi-Hiroshima, Hiroshima 739-0046, Japan.

[4]State Key Laboratory of Functional Materials for Informatics and Center for Excellence in Superconducting Electronics, Shanghai Institute of Microsystem and Information Technology, Chinese Academy of Sciences, Shanghai 200050, China.

[5]Neutron Science and Technology Center, Comprehensive Research Organization for Science and Society, Tokai, Ibaraki 319-1106, Japan

[6]Neutron Science Section, J-PARC Center, Japan Atomic Energy Agency, Ibaraki 319-1195, Japan

[7]Neutron Science Division, Oak Ridge National Laboratory, Oak Ridge, Tennessee 37831, USA

[8]Australian Nuclear Science and Technology Organisation, Locked bag 2001, Kirrawee DC, New South Wales 2232, Australia

[9]State Key Laboratory for Magnetism, Institute of Physics, Chinese Academy of Sciences, Beijing 100190, China

[†]These authors contributed equally to this work.

[*]Correspondence should be addressed to L.W. (wuls@sustech.edu.cn), C.C. (chency@sustech.edu.cn) and Q.L. (liuqh@sustech.edu.cn)

In a Dirac semimetal, the massless Dirac fermion has zero chirality, leading to surface states connected adiabatically to a topologically trivial surface state as well as vanishing anomalous Hall effect (AHE). Recently, it is predicted that in the nonrelativistic limit of certain collinear antiferromagnets, there exists a type of chiral "Dirac-like" fermion, whose dispersion manifests four-fold degenerate crossing points formed by spin-degenerate linear bands, with topologically protected Fermi arcs. Such unconventional chiral fermion, protected by a hidden SU(2) symmetry in the hierarchy of an enhanced crystallographic group, namely spin space group, is not experimentally verified yet. Here, by angle-resolved photoemission spectroscopy measurements, we reveal the surface origin of the electron pocket at the Fermi surface in collinear antiferromagnet $CoNb_3S_6$. Combining with neutron diffraction and first-principles calculations, we suggest a multidomain collinear AFM configuration, rendering the the existence of the Fermi-arc surface states induced by chiral Dirac-like fermions. Our work provides spectral evidence of the chiral Dirac-like fermion caused by particular spin symmetry in $CoNb_3S_6$, paving an avenue for exploring new emergent phenomena in antiferromagnets with unconventional quasiparticle excitations.

**Introduction**

The Dirac fields obey the famous Dirac equation, $(-i\alpha^i\partial_i + m\beta)\psi(x) = i\partial_0\psi(x)$, where $\alpha^i = \tau_x \otimes \sigma_i$ and $\beta = \tau_z \otimes \sigma_0$. With the operators furnishing a four-dimensional irreducible representation of the Lorentz group, the Dirac field can be decomposed into two two-component Weyl fields with opposite chirality in the limit of zero mass. There are several manifestations of the Dirac equation in condensed matter systems, such as the quasiparticle dispersion in graphene[1], topological insulators[2-4], Dirac semimetals[5-9], Weyl semimetals[10-12], and *d*-wave high-temperature superconductors[13]. In Dirac semimetals, the chirality of a massless Dirac fermion must be zero because the space-time *PT*-symmetry (*P* and *T* denote space inversion and time reversal, respectively) forces the two branches of each doubly degenerate band to have opposite Berry curvatures (Fig. 1a). Hence, the Fermi arc surface states connecting two Dirac points in a Dirac semimetal are generally not topologically protected, unlike the Fermi arc connecting chiral Weyl fermions[14]. On the other hand, chiral fermions with charge-2 chirality have been predicted and measured in materials such as CoSi[15-19], of which the band structure manifests four-fold degeneracy node protected by nonsymmorphic symmetry and nondegenerate bands off the high-symmetry point (Fig. 1a).

Recently, it is predicted that two Weyl fields with the same chirality could be connected together to form a "Dirac-like" fermion [20], which manifests four-fold degenerate nodes formed by two doubly degenerate bands while carrying Chern numbers $C = \pm 2$ (Fig. 1a). Interestingly, the symmetry that connects the two Weyl fields is a counterpart of isospin SU(2) symmetry that relates a proton and a neutron in high-energy physics (see Supplementary S1). In solid state physics, such continuous symmetry does not exist in the framework of conventional (magnetic) crystallographic groups. Instead, the generators of such hidden SU(2) symmetry belong to spin group, which involves partially decoupled spatial and spin operations[21-23], providing a symmetry description of magnetic materials with local moments in the non-relativistic limit. Despite several predicted material candidates, such chiral Dirac-like fermions is either not experimentally observed in quantum materials, or associated with any emergent phenomena.

In this work, we provide experimental evidence of the existence of such exotic fermions in an antiferromagnet CoNb$_3$S$_6$, which caught great interest due to its surprisingly large AHE[24-29]. By angle-resolved photoemission spectroscopy (ARPES) measurements, we reveal that the electron pockets of CoNb$_3$S$_6$ at the Fermi surface exhibit a two-dimensional nature. Combining with neutron diffraction and first-principles calculations, we suggest a multidomain collinear AFM configuration, rendering the the existence of the Fermi-arc surface states induced by chiral Dirac-like fermions. In addition, we discuss the effects of the chiral Dirac-like fermions as an essential element of the unexpected large anomalous Hall effect.

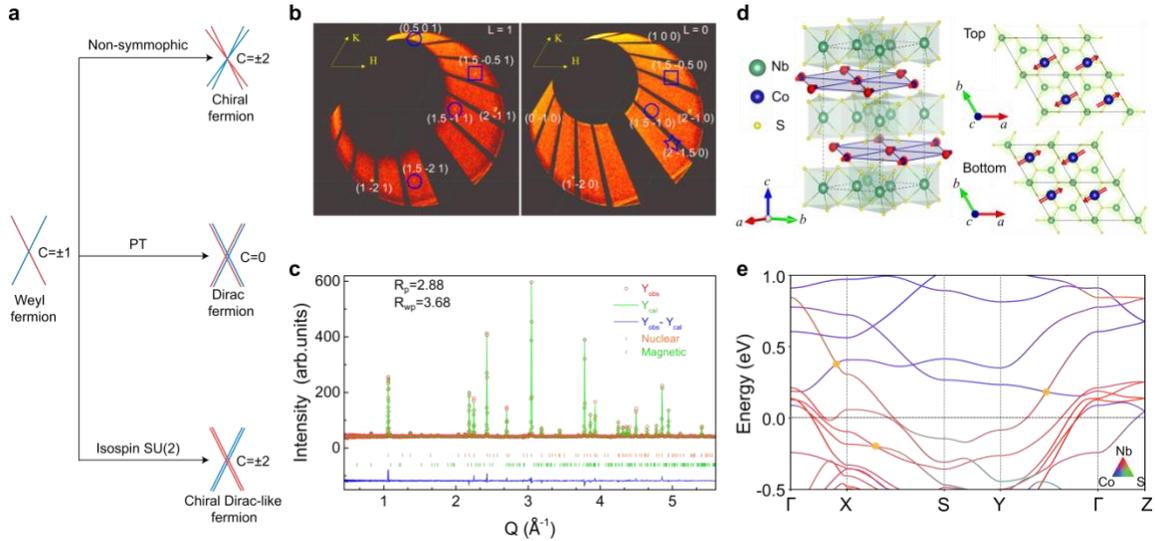

**Figure 1. Structure of the chiral Dirac-like fermion material candidate CoNb$_3$S$_6$. a**, Schematics of charge-1 Weyl fermion as a building block to compose charge-2 chiral fermion, Dirac fermion, and charge-2 chiral Dirac-like fermion. **b**, Single-crystal neutron diffraction image in the (H, K, L = 1) and (H, K, L = 0) scattering plane measured at 3 K. The reflections marked by blue circles, stars and squares are from three different magnetic domains indexed by three equivalent magnetic propagation vectors, $q_{m1}$=(0.5, 0, 0), $q_{m2}$=(0, 0.5, 0), and $q_{m3}$=(0.5, -0.5, 0), respectively. **c**, Powder neutron diffraction pattern with Rietveld refinement fit at 10 K. **d**, The AFM magnetic structure of CoNb$_3$S$_6$ revealed by the Rietveld refinement of powder neutron diffraction patterns. Note that due to the effect of the six magnetic $M$ domains induced by the site point group $D_3$ of Co, the specific

moment direction in the *ab* plane cannot be exactly determined.  **e**, DFT-calculated band structure of $CoNb_3S_6$ with the experimentally measured AFM order without SOC.  Yellow circles indicate the positions of the chiral Dirac points.

**Magnetic structure of $CoNb_3S_6$**

Our magnetization measurements suggest a phase transition around $T = 28.3$ K, with most of the moments ordered antiferromagnetically in the *ab* plane, and a weak ferromagnetic component along the *c* axis. This observation is consistent with the previous reports[24-27]. The detailed magnetization data is presented in Supplementary S2, S3. Fig. 1b presents the single-crystal neutron diffraction patterns of $CoNb_3S_6$ in the (H, K, 1) and (H, K, 0) scattering plane at 3 K. The magnetic peaks marked by the blue circles, stars and squares can be indexed by three different magnetic wave vectors (0.5, 0, 0), (0, 0.5, 0) and (0.5, -0.5, 0), respectively, indicating three types of magnetic domains rotated 120° from each other. Due to the limited magnetic reflections of the single-crystal diffraction experiment, an additional powder neutron diffraction experiment was performed at 10 K < $T_N$ to determine the magnetic structure of $CoNb_3S_6$, as shown in Fig. 1c. Consistent with the single-crystal diffraction results, extra weak magnetic reflections are observed and indexed by the same magnetic propagation vector $q_m = (0.5, 0, 0), (0, 0.5, 0)$ or $(0.5, -0.5, 0)$. Representation analysis was applied to analyze the possible magnetic structures[30]. For the space group $P6_322$ with Co site at (1/3, 2/3, 1/4) and $q_m = (0.5, 0, 0)$, the spin configuration can be described by four different irreducible representations. By Rietveld refinement, we found that $\Gamma_4$ could give the best fit with $R_p = 2.88$ and $R_{wp} = 3.68$ for powder neutron diffraction data. The resulting magnetic moment on the site (1/3, 2/3, 1/4) are antiferromagnetically coupled to the site (2/3, 1/3, 3/4). As schematically presented in Fig. 1d, the refined magnetic structure of $CoNb_3S_6$ shows a collinear magnetic configuration[31]. For the intralayer, the local moments on the neighboring Co sites are antiferromagnetically coupled along the *a* axis, but ferromagnetically coupled along the *b* axis. By the refinement, we find the Co moments are lying in the *ab* plane and the ordered moment of $Co^{2+}$ is about 1.64(7) $\mu_B$/Co. However, due to the $D_3$ site point group of $Co^{2+}$ ions, the in-plane spin orientation cannot be distinguished. The detailed analysis of the magnetic structure can be found in Supplementary S2.

**Emergence of Fermi-arc surface states**

We perform DFT band structure calculation of bulk $CoNb_3S_6$ based on the measured AFM order, as shown in Fig. 1e. Since the bands of interest are dominated by Co 3*d* orbitals with weak SOC, we ignore SOC for the calculations that compare with the ARPES measurement and leave the SOC effects in later discussions. There are two main features in the calculated band structure: Firstly, although *PT* symmetry is absent, the energy bands of any momenta are doubly degenerate. Such degeneracy is unique in magnetic materials without SOC in that it is protected by the so-called spin space group symmetry, which involves independent spin and spatial rotations compared with the conventional magnetic space group[21,23]. In $CoNb_3S_6$, the collinear AFM order guarantees U(1) symmetry along the $x$ axis $\{U_x(\theta)||E|0\}$ and a 180° pure spin rotation along the $z$ axis followed by a fractional translation $\{U_z(\pi)||E|\,\tau_{(a+b)/2}\}$, ensuring doubly degenerate bands throughout the Brillouin zone (see Supplementary S4.1). Secondly, the band crossings are all four-fold degenerate Dirac-like points, which could appear either at arbitrary momenta or high-symmetry lines. There are multiple four-fold Dirac-like points around the calculated Fermi level and some located along $\Gamma - X$ and $\Gamma - Y$ lines (~ 0.4 eV above the Fermi level). Interestingly, unlike the degeneracy protected by *PT*, all the Dirac-like points manifest Chern numbers $C = \pm 2$ rather than 0, manifesting robust Fermi-arc surface states (see Supplementary S4.2).

We next perform ARPES measurements on the natural cleavage plane (*ab* plane) to directly visualize the band structure of $CoNb_3S_6$. Although the AFM order enlarges the unit cell leading to a rectangular Brillouin zone (BZ), the ARPES measured spectral intensity exhibits a hexagonal symmetry matching the nonmagnetic 3D BZ (Fig. 2a). This comes from the fact that ARPES spectral intensity averages photoelectrons excited from energy degenerate AFM domains with three different orientations as revealed by the neutron diffraction results. Thus, we use the nonmagnetic 3D BZ to describe the ARPES data measured at $T = 8\ K$, *i.e.*, the AFM phase. The general band structure along the high-symmetry line $\bar{\Gamma} - \bar{K} - \bar{M} - \bar{\Gamma}$ is shown in Fig. 2b. Close to the Fermi level, the ARPES spectra is dominated by a hole-like, highly dispersed band (labeled as $\alpha$) centered at $\bar{\Gamma}$ and an electron-like, shallow band (labeled as $\beta$) centered at $\bar{K}$. To define the precise position of high-symmetry points in the 3D BZ and uncover the $k_z$ dependence of these bands, we perform photon-energy dependent measurements ($hv = 60 \sim 165\ eV$), with the momentum cut fixed along $\bar{K} - \bar{\Gamma} - \bar{K}$ direction (Figs. 2d and 2e). As shown in Fig. 2d by the Fermi surface mapping in $k_z - k_x$ plane, both $\alpha$ and $\beta$ bands show no observable dispersion with $k_z$, despite some intensity change, consistent with its layered lattice structure. The $k_z$ periodicity can only be observed if we choose the energy window $\sim 1\ eV$ below the Fermi level and focus on the spectral intensity variation from $\bar{\Gamma}$. As shown in Fig. 2c, broad but alternating electron-like and hole-like features can be distinguished as indicated by the superposed white dotted lines. It is noted that the $k_z$ dispersion shows a $4\pi/c$ periodicity with lattice constant $c = 11.886$ Å, because each nonmagnetic unit cell contains two -NbS$_2$-Co$_{1/3}$-units.

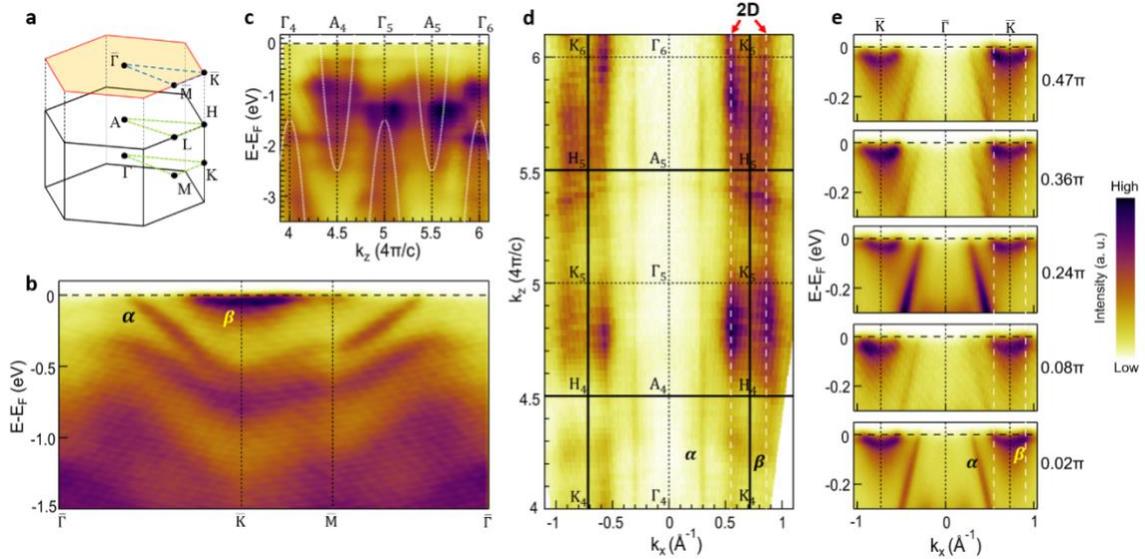

**Figure 2. Band structure and its $k_z$ dependence of CoNb$_3$S$_6$ measured by ARPES at $T = 8\ K$. a,** 3D bulk and 2D surface Brillouin zone of CoNb$_3$S$_6$. **b,** Band spectrum along high-symmetry lines $\bar{\Gamma} - \bar{K} - \bar{M} - \bar{\Gamma}$. The hole pocket that is centered at $\bar{\Gamma}$ and crosses Fermi level is labeled as $\alpha$. The electron pocket centered at $\bar{K}$ is labeled as $\beta$. **c,** Spectral

intensity along $k_z$ the direction taken with photon energies ranging from 60 to 165 eV, superposed with periodic dispersion (white dotted lines). **d**, Fermi surface in $k_x - k_z$ plane obtained by photon-energy dependent ARPES measurement. The white dashed lines in **d** indicate the $k_z$ independence of $\beta$ band. **e**, Band spectra along $\overline{K} - \overline{\Gamma} - \overline{K}$ from various $k_z$ values. White dashed lines indicate the same $k_F$ of $\beta$ at different $k_z$.

The 2D nature of $\alpha$ and $\beta$ bands are further elaborated by examining their dispersion at different $k_z$ values. As shown in Fig. 2e, we plot $\overline{K} - \overline{\Gamma} - \overline{K}$ cuts from five randomly selected $k_z$ values, all of which show almost the same dispersion for both $\alpha$ and $\beta$ bands. In particular, the Fermi momentum ($k_F$) of $\beta$ band is indicated by the white dashed lines, and it remains constant with $k_z$, strongly demonstrating its 2D nature. Previously, this electron pocket was attributed to the bulk electronic structure and are dominated by Co atoms[28,29], evidenced by the $k_z$ dispersion observed by ARPES using soft X-ray photons[27]. Here we use UV photons with much higher energy and momentum resolution to clearly prove its $k_z$ independence and will discuss its surface origin in the following.

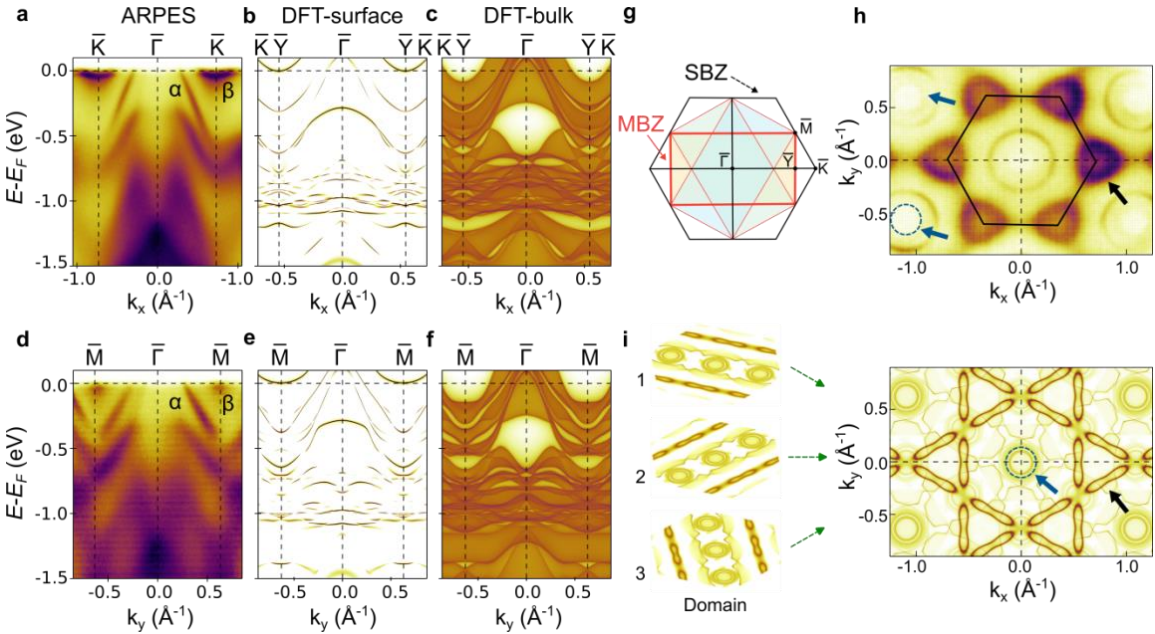

**Figure 3. Fermi-arc surface states of CoNb$_3$S$_6$. a** and **d** ARPES spectra taken with 120 eV photons along the $\overline{K} - \overline{\Gamma} - \overline{K}$ and $\overline{M} - \overline{\Gamma} - \overline{M}$, respectively. **b, c, e, f,** Projection of DFT calculated bulk and surface bands along $\overline{K} - \overline{Y} - \overline{\Gamma} - \overline{Y} - \overline{K}$ and $\overline{M} - \overline{\Gamma} - \overline{M}$, respectively. **g**, Geometric relation between nonmagnetic surface BZ (SBZ, black solid lines) and antiferromagnetic BZ (MBZ, colored solid lines). **h**, ARPES Fermi surface mapping with SBZ plotted. **i**, Left: DFT calculated Fermi surfaces with only surface state spectral weight for the three equivalent domains. Right: superposition of the three differently oriented Fermi surfaces to construct the experimental Fermi surface with sixfold rotation symmetry.

We then compare the ARPES spectra to the projection of DFT calculated bulk and surface bands to fully demonstrate the surface origin of $\beta$ band and its association with the predicted chiral Dirac-like fermions. Since the structure of $CoNb_3S_6$ is indeed stacking $NbS_2$ layers with a Co-layer intercalation, the calculated surface states of $NbS_2$ termination are adopted for comparison. We find that almost all the ARPES measured low-energy band features (including $\alpha$ and $\beta$ bands, Figs. 3a, 3d) can be reproduced by DFT projected surface (Figs. 3b, 3e) and bulk calculations (Figs. 3c, 3f). Fig. 3a shows the ARPES spectra along $\bar{K} - \bar{\Gamma} - \bar{K}$. The $\beta$ band centered at $\bar{K}$ comes from an electron pocket with its band bottom slightly below the Fermi level. Such feature can be well reproduced by the DFT-calculated surface states as shown in Fig. 3b. The ARPES spectra along $\bar{M} - \bar{\Gamma} - \bar{M}$ also reveals weak spectral weight centered at $\bar{M}$ (Fig. 3d). We attribute this feature to the tail of the $\beta$ surface band which locates slightly above the Fermi level at $\bar{M}$ as shown in Fig. 3e. Such a tail is also visible for the $\beta$ band at $\bar{K}$, likely from the incoherent electron scattering off other entities such as disorders, bosons and so on[32]. In Supplementary S4.2, orbital projection analysis shows that $\beta$ band is dominated by intercalated Co-3$d$ atoms. Further calculations show that the $\beta$ pocket are indeed the Fermi-arc surface states originated from the predicted chiral Dirac-like points located 0.17 eV above the Fermi level, rather than a trivial surface resonance of the bulk band (see Fig. S9a). Therefore, the agreement between our photon-energy dependent ARPES measurement and DFT calculation supports a surface origin of the electron pockets.

While along $\bar{M} - \bar{\Gamma} - \bar{M}$ the ARPES and DFT surface spectra show agreement, there is a slight mismatch of the band edge (minimum) of the $\beta$ pocket at $\bar{K}$ obtained from ARPES and DFT. Such discrepancy is because our DFT calculations is based on a single-domain collinear AFM order, with the rectangular magnetic BZ (MBZ) shown in Fig. 3g. As a result, the band edge appears at the $\bar{Y}$ point, the boundary of the MBZ. On the other hand, the APRES spectra inevitably averages multiple magnetic domains with the same ground-state energy, thus restoring the hexagonal symmetry for the surface BZ (SBZ) and band edges appearing at the $\overline{K}$ valley ($\overline{\Gamma K} = \frac{4}{3}\overline{\Gamma Y}$). Fig. 3h shows the measured Fermi surface indicating an identical shape and size of BZ to that of a nonmagnetic unit cell. However, the measured Fermi surface as well as the dispersion can hardly be reproduced by the nonmagnetic calculation of $CoNb_3S_6$ (see Supplementary S4.3), while the AFM order gives rise to a rectangular MBZ with lower symmetry.

To solve the dilemma, we consider three equivalent $q$-vectors related by $C_3$ rotational symmetry, rendering three energetically degenerate magnetic domains with three rectangular MBZs rotating 120 degrees with respect to each other. Therefore, an effective hexagonal SBZ is formed, with the size identical to the nonmagnetic one (Fig. 3g). The left panel of Fig. 3i shows the calculated surface state Fermi surfaces for each single AFM domain and illustrates the formation of the hexagonal Fermi surface by superposing the Fermi surfaces of these three equivalent domains. Figs. 3h and the right panel of Fig. 3i compare the Fermi surfaces from ARPES and DFT. ARPES mapping reveals six triangular pockets centered at $\overline{K}$ formed by the $\beta$ pocket (indicated by the black arrow in Fig. 3h). Instead of an open line, the closed shape of such pocket is in line with the fact that the bottom of $\beta$ pocket resides below the Fermi level at $\overline{K}$ and above it at $\overline{M}$ (see Supplementary S4.2). According to our DFT calculation, each triangular pocket is formed by three peanut-like surface pockets from three equivalent AFM domains. The broadness of the measured surface bands, as evidenced by the momentum distribution curve analysis in Supplementary S5, may smear out the fine structure of the calculated surface states, resulting into broad $\beta$ band features centered at $\overline{K}$. In addition, the small pocket centered at $\overline{\Gamma}$ of the second BZ (Fig. 3h, blue arrow and blue dashed circle) is in consistent with our DFT calculation (Fig. 3i, blue arrow), further validating the general agreement. We note that due to the dipole matrix element effect in photoemission experiments (also see Supplementary S6), such small pocket is not visible in the first BZ. The general agreement between ARPES and DFT results throughout this work validates the above arguments and the existence of Fermi arc surface states associated to the chiral Dirac-like fermions in $CoNb_3S_6$.

**Anomalous Hall effect**

The most intriguing finding in CoNb$_3$S$_6$ system is the emergence of a substantial anomalous Hall effect, accompanied by a small but not negligible net magnetic moment[24-27,33]. These results are also verified by our transport measurements. Fig. 4a presents the field evolution of the Hall resistivity measured from 22 K to 29 K with I // $a$ and B // $c$. For $T$ = 29 K > $T_N$, linear dependence of the Hall resistivity as a function of magnetic field (brown line in Fig. 4a) was observed. The positive slope of the Hall resistivity suggests that holes are the majority charge carriers in CoNb$_3$S$_6$. When the temperature is below 23 K, the coercive field becomes larger than 14 T. By subtracting the linear ordinary Hall background and using $\sigma_{xy}^A = \rho_{xy}^A/((\rho_{xy}^A)^2 + (\rho_{xx})^2)$, a large anomalous Hall conductivity $\sigma_{xy}^A \sim 92\ (\Omega \cdot \text{cm})^{-1}$ was obtained at 26 K (Fig. 4b). More detailed information is presented in Supplementary S7. To examine the ferromagnetic contribution to the anomalous Hall conductivity, the field dependent ferromagnetic component (- $\Delta M$) along the $c$ axis is plotted as well (orange empty circles in Fig. 4b). The measured $\Delta M$ is ~ 0.001 $\mu_B$/Co, which seems too insignificant to induce a faily large AHE. Such a strong scaling between the AHE and ferromagnetic canting $\Delta M$ could be explained by the large hidden Berry curvature due to the chiral Dirac-like fermions.

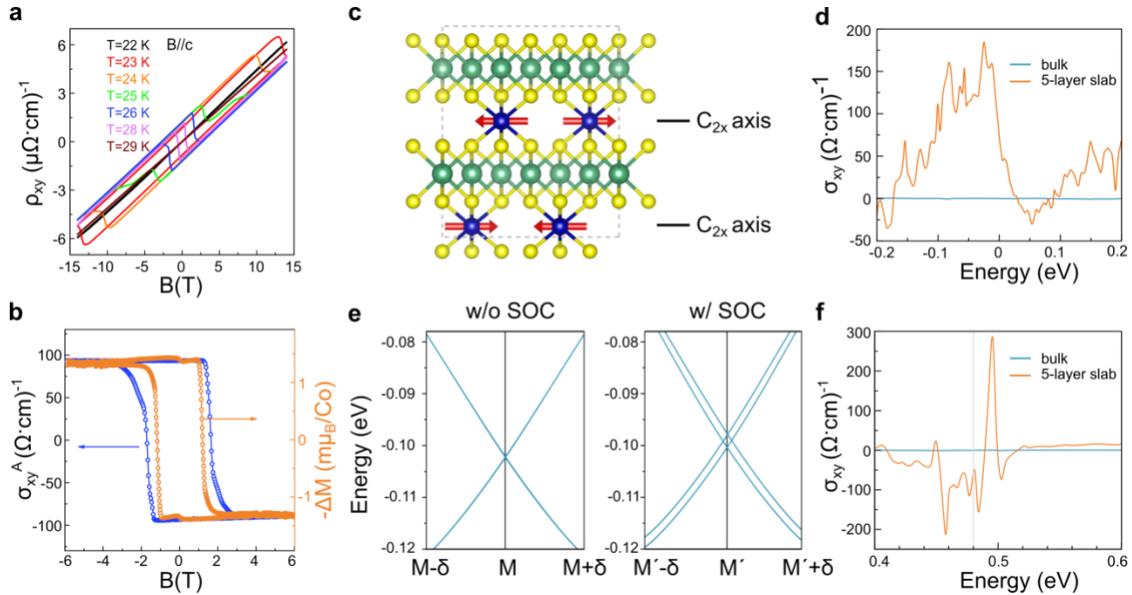

**Figure 4. Origin of the large anomalous Hall in CoNb$_3$S$_6$. a,** Field evolution of the Hall resistivity measured at different temperatures. **b,** The anomalous Hall conductivity (blue empty circles) scales with the ferromagnetic component (orange empty circles) at $T = 26$ K. The magnetic field is along the $c$ axis. **c,** Crystal structure of CoNb$_3$S$_6$ where the black ticks show the plane to which rotation axis of $C_{2x}$ exists. **d,** Anomalous Hall conductivity near the Fermi level. **e,** Band dispersion near a chiral Dirac point of CoNb$_3$S$_6$ without and with SOC, where $M = (0.4998, 0.1495, -0.0003)$, $M' = (0.4946, 0.1474, -0.0001)$, $\delta = (0.1000, 0.0000, 0.0000)$. Anomalous Hall conductivity $\sigma_{xy}$ of Bulk and 5-layer slab calculated near the energy level of chiral Dirac points shown in **f**, where the grey dashed line shows the energy level of chiral Dirac point (0.48 eV).

Due to the symmetry operation of time-reversal combined with nonsymmorphic translation $\{T||E|\tau_{(a+b)/2}\}$ (see Supplementary S4.1, Table S2), bulk CoNb$_3$S$_6$ cannot exhibit finite anomalous Hall conductivity. However, rather than being intrinsically absent, the Berry curvature originated from the nontrivial bands are large yet compensated by the global high symmetry[34,35]. Therefore, the small ferromagnetic (FM) canting along the $z$-axis and finite SOC play a role of symmetry breaking that reveals the large Berry curvature effect hidden in the otherwise doubly degenerate bands, thus leading to finite anomalous Hall effect. Here we consider bulk and 5-layer thin film CoNb$_3$S$_6$ with the experimental AFM order. Our calculations show that, although under small magnetic canting, bulk states exhibit anomalous Hall conductivity (AHC) smaller than 1 $(cm \cdot \Omega)^{-1}$, large AHC approaching 185 $(cm \cdot \Omega)^{-1}$ emerges for the thin film, when the chemical potential is 0.01-0.03 eV below the theoretical Fermi level (Fig. 4d). The remarkable difference between bulk and thin film could be attributed to the fact that all the rotation symmetries in thin film CoNb$_3$S$_6$ are broken even without small magnetic canting (see Supplementary S8.1, Table S3), as is shown in the crystal structure of CoNb$_3$S$_6$ (Fig. 4c).

There are quite a few chiral Dirac-like fermions near the Fermi level, when considering SOC, some of them split to a twin pair of conventional Weyl points with identical chirality (Fig. 4e) and others are gapped. Since chiral Dirac points is obstructed by the complicated generic metallic bands around the Fermi level (Fig. 1e). To reveal the relationship of the chiral Dirac-like fermions and the AHC more clearly, we show another energy window (0.4 ~ 0.6 eV) where there is only one pair of chiral Dirac-like points at 0.48 eV. With SOC and FM canting, the pair of chiral fermions are gapped, leading to sharp peaks of AHC (~200 $(cm \cdot \Omega)^{-1}$) around the corresponding chemical potential (Fig. 4f). Therefore, the large AHE in collinear AFM $CoNb_3S_6$ results from the Berry curvature of the chiral Dirac-like fermions as well as symmetry breaking by SOC, ferromagnetic canting. Furthermore, our results are also consistent with the fact that AHC measured in thin films is much larger than that measured in thick slabs[25].

**Discussion**

Except for mechanism discussed above, due to the complexity of magnetic configuration and surface topography in $CoNb_3S_6$, other mechanisms cannot be fully ruled out. In this section, we discuss our findings in connection with recent progress in this particular material system, to offer a more comprehensive description reconciling observations by different experimental techniques. Notably, motivated by the anomalous transports, a triple-**q** AFM order has been recently reported in $CoM_3S_6$ (M = Ta and Nb) by using polarized neutron scattering measurements[36,37], suggesting a "spontaneous topological Hall effect", i.e., sizable AHE originated from the noncoplanar magnetic order rather than the net moment (see Supplementary S8.2). However, the observed transport property of $CoNb_3S_6$ exhibits remarkable sample dependency, indicating that the existence of such a tiny net moment indeed plays a crucial role. By measuring different $Co_xNb_3S_6$ samples with slightly changed Co composition (0.92< $x$ < 1), we find the correlation between the existence of net magnetization and the existence of AHE (see Supplementary S9). These results are also confirmed by the recent measurements with varying-composition smaples[26,27]. Nevertheless, the absence of AHE without net magnetization in $CoNb_3S_6$ cannot be reconciled by the triple-**q** scenario. In addition, according to our DFT calculation, the Fermi surface of triple-**q** configuration cannot well reproduce the ARPES-measured Fermi surface (see Supplementary S10).

From the aspect of neutron diffraction, the direct evidence to distinguish triple-**q** magnetization and multidomain single-**q** configuration is still lacking. The existing zero-field polarized neutron scattering taken by H. Takagi et al.[36] unfortunately cannot provide enough information because the two magnetic structures share similar diffraction patterns and peak intensities[25]. Similar phenomena are also observed in the $Na_2Co_2TeO_6$, thereby posing challenges in distinguishing between its zigzag order and triple-**q** order through zero-field neutron scattering[38,39]. To overcome this, high-field neutron scattering experiments are desirable to distinguish these possibilities[38,39]. However, due to the presence of a large antiferromagnetic exchange interaction (the fitted Curie-Weiss temperature is -174 K) in the $CoNb_3S_6$ system, a very high magnetic field (~ 60 T) is required to fully polarize the magnetic moments. Overall, considering all the experimental results mentioned above, the complexity of $CoNb_3S_6$ family exhibiting interesting sample dependence may imply a composited phase diagram with distinct magnetic configurations, calling for more direct evidence for future studies.

Last but not the least, we discuss the impact of our work regarding the classification and related properties of collinear AFM materials in zero SOC limit, which have caught remarkable attention recently. The predominant example is the emerging field of altermagnetism[40,41], with spin splitting band structures originated from the AFM order. By definition, altermagnets are a special type of AFM where the sublattices with opposite magnetic moments are symmetry connected by operations other than inversion or translation; otherwise, there is two-fold spin degeneracy throughout the Brillouin zone. On the other hand, our work reveals that for spin-degenerate AFM, there is significant difference between the two classes, i.e, the critical symmetry connecting different sublattices is inversion, or translation[42]. The former corresponds to the conventional AFM with *PT* symmetry, manifesting zero topological charge for a conventional Dirac semimetal. For the latter, despite spin degeneracy, the band nodes manifest nonzero topological charges (e.g., due to $\{U_z(\pi)||E|\,\boldsymbol{\tau}_{(\boldsymbol{a}+\boldsymbol{b})/2}\}$ in $CoNb_3S_6$), corresponding to chiral Dirac semimetal. Therefore, our work reveals unconventional properties in a class of spin-degenerate AFM, paving an avenue for exploring unexpected emergent phenomena for AFM spintronics.

**Methods**

**Sample growth**

Single crystals of CoNb$_3$S$_6$ were grown by chemical vapor transport (CVT) using iodine as the transport agent. Cobalt powder (99.998%), niobium wire (99.995%) and sulfur powder (99.998%) in 1 : 3 : 6 molar ratio were loaded together with 0.3 g of iodine in a silica tube of 17 mm inner diameter and 150 mm length. After evacuated down to 10$^{-3}$ Pa and sealed, the tube was placed in a two-zone horizontal tube furnace, and the source and growth zones were raised to 1123 K and 1023 K, and then held for 7 days. The hexagonal black crystals with lateral dimensions up to several millimeters can be obtained.

**Magnetization and Electric Transport**

The magnetization measurements were performed on a Quantum Design Magnetic Property Measurement System (MPMS) with a single crystal mounted on a quartz stick. The electric transport measurements were performed on a 14 T Quantum Design Physical Property Measurements System (PPMS) with a conventional 4-probe method. Al wires of 25 μm diameter were attached to the sample by using a wire bonding machine. An electric current of 2 mA was applied along the *a* axis, and the magnetic field was oriented along the *c* axis.

**Neutron diffraction**

The single crystal neutron diffraction experiments were carried out using a single-crystal neutron Laue diffractometer, KOALA[43], at the OPAL reactor at ANSTO and a time-of-flight single-crystal neutron Laue diffractometer, SENJU[44], at the Materials and Life Science Experimental Facility of the Japan Proton Accelerator Research Complex. Two different single crystals (one with AHE, the other without AHE) with dimensions of about 3 × 3 × 0.2 mm$^3$ were used for the neutron diffraction experiments and the diffraction patterns were both collected at 3 K (below $T_N$) and 40 K (above $T_N$), respectively. The powder neutron diffraction experiment was performed on the time-of-flight powder diffractometer, POWGEN[45], at the Spallation Neutron Source at Oak Ridge National Laboratory. The powder sample with a total mass of ~ 1g was prepared by grinding about hundreds of single crystals and the powder neutron diffraction data were acquired between 10 K and 300 K using the 0.8 and 2.67 Å instrumental configurations. All of the neutron diffraction data were analyzed using the Rietveld refinement program FULLPROF suite[46].

**Angle-resolved Photoemission Spectroscopy (ARPES) Experiments**

The ARPES measurement was taken with 60~165 eV photons and Scienta Omicron DA30 analyzer. The sample was cleaved in ultra-high vacuum with pressure lower than $1 \times 10^{-10}\ torr$. During the measurement, the temperature of the sample was kept at around $8\ K$. The beam spot of the light is less than 50 $\mu m$.

**First-principles calculations**

The first-principles calculations were carried out using projector-augmented-wave (PAW) method[47], implemented in Vienna ab initio simulation package (VASP)[48] within the framework of density-functional theory[49,50]. The exchange and correlation effects were accounted by the generalized gradient approximation (GGA) with the Perdew-Burke-Ernzerhof (PBE) formalism[51]. An energy cut off of 520 eV is used in our calculations. The whole Brillouin-zone was sampled by $5 \times 8 \times 4$ Monkhorst-Pack grid[52] for all cells. Due to the local magnetic moments contributed from 3$d$ electrons in Co atoms, GGA+U approach[53] within the Dudarev scheme[54] is applied and we set the U on Co to be 1 eV, which produces local magnetic moments of 1.8 $\mu_B$ consisting well with the experiments[31]. A tight-binding Hamiltonian is obtained base on maximally localized Wannier functions[55,56] of Co-3$d$ , Nb-4$d$, S-3$p$ orbitals, from which the topological surface states and Chern number are calculated. The iterative Green's function implemented in WannierTools package is used for surface states calculations[57].

**Acknowledgements**

We thank Prof. Jia-Xin Yin for helpful discussions. This work was supported by National Key R&D Program of China under Grant Nos. 2020YFA0308900 and 2022YFA1403700, the National Natural Science Foundation of China (Grant Nos.12074163, No.12134020, No.11974157, No.12104255, No.12004159, 1237041347 and 12104255), Guangdong Provincial Key Laboratory for Computational Science and

Material Design under Grant No. 2019B030301001, the Science, Technology and Innovation Commission of Shenzhen Municipality (Nos. ZDSYS20190902092905285 and KQTD20190929173815000), Guangdong Basic and Applied Basic Research Foundation (Grant Nos. 2022B1515020046, 2021B1515130007, 2022A1515011915, 2019A1515110712 and 2021B1515130005), Shenzhen Science and Technology Program (Grant No. RCJC20221008092722009 and RCBS20210706092218039), the Guangdong Innovative and Entrepreneurial Research Team Program (Grant No. 2019ZT08C044), Center for Computational Science and Engineering and Core Research Facilities of Southern University of Science and Technology. The authors would also like to acknowledge the beam time awarded by Australia's Nuclear Science and Technology Organisation (ANSTO) through proposal No. P8130. The neutron experiment at the Materials and Life Science Experimental Facility of the Japan Proton Accelerator Research Complex (J-PARC) was performed under a user program (Proposal No. 2019B0140). The ARPES measurement was performed at the Hiroshima Synchrotron Radiation Center (HiSOR) of Japan under proposal number 22BG023 and 22BG029, and Shanghai Synchrotron Radiation Facility (SSRF) BL03U under proposal number 2022-SSRF-PT-020848. The authors acknowledge the assistance of SUSTech Core Research Facilities.